\documentclass[12pt,notitlepage]{article}
\usepackage{eucal,amsmath,amsfonts,amsthm,amssymb,latexsym}

\theoremstyle{plain}

\newtheorem{thm}{Theorem}

\newtheorem{Lma}{Lemma}

\theoremstyle{definition}

\newcommand{\RR}{\mathbb{R}}

\newcommand{\rd}{\mathrm{d}}
\newcommand{\rw}{\mathrm{w}}

\newcommand{\ww}{\mathsf{\omega}}

\newcommand{\lp}{\overline{\lim}}
\newcommand{\lf}{\underline{\lim}}

\newcommand{\lindent}
{
\setlength{\labelwidth}{2cm}\setlength{\leftmargin}{2.0cm}
\setlength{\labelsep}{0.5cm}\setlength{\rightmargin}{1.0cm}
\setlength{\parsep}{0.5ex plus 0.2ex minus0.1ex}
\setlength{\itemsep}{0ex plus 0.2ex}
}

\newcommand{\findent}
{
\setlength{\labelwidth}{2cm}\setlength{\leftmargin}{1.7cm}
\setlength{\labelsep}{0.5cm}\setlength{\rightmargin}{1.0cm}
\setlength{\parsep}{0.5ex plus 0.2ex minus0.1ex}
\setlength{\itemsep}{0ex plus 0.2ex}
}

\newcounter{fact}
\newcounter{prop}

\begin{document}
\setcounter{page}{1}

\title{Near Field Behavior of Static Spherically Symmetric Solutions of Einstein SU(2) - Yang Mills Equations \thanks{Part of a doctoral thesis at the University of Michigan under the supervision of Professor Joel A. Smoller}}
\author{Alexander N. Linden\thanks{Zorn Assistant Professor, Indiana University}}
\date{}
\maketitle


\begin{abstract}
We consider static spherically symmetric solutions of the Einstein equations with cosmological constant $\Lambda$ coupled to the SU(2) Yang Mills equations.  We prove that under relatively mild conditions, any solution can be continued back to the origin of spherical symmetry and that the qualitative behavior of the solutions near the origin does not depend on $\Lambda$.  

\end{abstract}
\begin{sloppypar}
\section{Background}
\label{back}
\setcounter{equation}{0}
\subsection{Introduction}
\label{intro}
$\;$

We consider the forces of gravity as modeled by Einstein's equations with cosmological constant $\Lambda$ coupled to SU(2)-Yang Mills fields.  In particular, we examine static spherically symmetric solutions.  This model gives rise to a system of ordinary nonlinear differential equations.  There have been numerous investigations of this system without the cosmological constant (see references).  In particular, it has been proved that there exists a one parameter family of solutions that are smooth at the center of symmetry.  This family also exists when $\Lambda\ne 0$.  There are also a class of solutions which are qualitatively like the Schwarzschild solution  and a class of solutions that behave like the Reissner-Nordstr\"om solutions (\cite{jS97}).  In this paper, we prove that regardless of the value of the cosmological constant, every solution is one of these three types.
   
The static spherically symmetric Einstein-Yang-Mills equations with cosmological constant take the form of two differential equations for the variables $A(r)$ and
$\rw(r)$:
\begin{equation}\label{Aeq}
rA'+2A{\rw'}^2=\Phi\;\mathrm{and}
\end{equation}
\begin{equation}\label{weq}
r^2A\rw''+r\Phi\rw'+\rw(1-\rw^2)=0
\end{equation}
where 
\begin{equation}\label{phidef}
\Phi=1-A-\frac{(1-\rw^2)^2}{r^2}-\Lambda r^2.
\end{equation}
$A$ is the same $A$ that appears in a spherically symmetric metric written as
\begin{equation}\label{metric}
\rd s^2 = -C^2(r,t)A(r,t)\; \rd t^2 + \frac{1}{A(r,t)} \;\rd r^2+r^2 \rd \Omega^2
\end{equation}
\[(\rd \Omega^2=\rd \phi^2+\sin^2\phi\; \rd \theta^2)\]
and w is the same w that appears in the spherically symmetric connection on an
\textbf{SU}(2) bundle; namely,
\begin{eqnarray}\label{connection}
\ww&=& \mathrm{a}(r,t) \mathbf{\tau}_3 \;\rd t + \mathrm{b}(r,t)\mathbf{\tau}_3 \;\rd r +\rw(r,t)\mathbf{\tau}_2
\;\rd \phi\nonumber\\
&+&(\cos\phi\mathbf{\tau}_3-\rw(r,t) \sin \phi
\mathbf{\tau}_1)\;\rd \theta.
\end{eqnarray}
$\mathbf{\tau}_i$ are the following matrices which form a basis of $su(2)$:
\[\mathbf{\tau}_1=i/2\left[
\begin{array}{cc}
0 & -1\\
-1 & 0
\end{array}
\right],\;
\mathbf{\tau}_2=i/2\left[
\begin{array}{cc}
0 & i\\
-i & 0
\end{array}
\right],\;
\mathbf{\tau}_3=i/2\left[
\begin{array}{cc}
-1 & 0\\
 0 & 1
\end{array}
\right].
\]

There is also an equation for $C$,
\begin{equation}\label{Ceq}
rC'=2{\rw'}^2C.
\end{equation}
However, equation~(\ref{Ceq}) separates from equations~(\ref{Aeq}) and (\ref{weq}) and yields
\begin{equation}\label{Csolved}
C(r)=C_0e^{\int_{t=0}^r(2{\rw'}^2/s)\;\rd s}
\end{equation}
where $C_0$ can be assigned arbitrarily.

We begin with any point $(\bar r,\bar A,\bar\rw,\rw')$ in $\RR^4$ with $\bar r$ and $\bar A$ both positive and consider the unique solution of equations~(\ref{Aeq}) and (\ref{weq}) that satisfies $(A(\bar r),\rw(\bar r),\rw'(\bar r))=(\bar A,\bar \rw,\bar\rw')$.  In Theorem~\ref{Main1} we prove that this solution can be continued back to the origin provided that $A$ remains positive on a sequence that approaches the origin.  Theorem~\ref{Mainpart} states that such solution is a member of the one parameter family of solutions whenever $A$ is positive and bounded.  Theorem~\ref{A>1} states that any other solution for which $A$ is positive behaves like a Reissner-Nordstr\"om solution.

The result of Therorem~\ref{Mainpart} has been proved in the case $\Lambda=0$ in~\cite{jS951} under slightly weaker hypotheses.   However, the arguments used there are not valid if $\Lambda\ne 0$. 
\subsection{Preliminaries}
\label{prel}
\setcounter{equation}{0}
$\;$

We begin by stating some basic facts regarding solutions to Equations~(\ref{Aeq}) and (\ref{weq}).
\begin{list}{\textit Fact \arabic{fact}:}
{\usecounter{fact}
\findent}
\item \label{reflect}The equations are invariant under the transformation $(r,A,\rw)\rightarrow (r,A,-\rw)$.
\end{list}
\begin{list}{\textit Fact \arabic{fact}:}
{\usecounter{fact}
\addtocounter{fact}{1}
\findent}
\item \label{wconst} If $\rw$ is constant, equation~(\ref{weq}) implies $\rw\equiv \pm 1$ or $\rw\equiv 0$.
\end{list}
\textbf{Proof of \textit{Fact~\ref{wconst}}}:  Integrating equation~(\ref{Aeq}) with $\rw^2\equiv 1$ yields
\begin{equation}\label{AdeSit}
A=1-\frac{2M}{r}-\frac{\Lambda r^2}{3}
\end{equation}
where $M$ is an arbitrary constant.  With a possibly rescaled $t$, this is a deSitter space  with constant Yang Mills connection,
\begin{equation}\label{deSitter}
\rd s^2=-(1-\frac{2M}{r}-\frac{\Lambda r^2}{3})\rd t^2+(1-\frac{2M}{r}-\frac{\Lambda r^2}{3})^{-1}\rd r^2+\rd \Omega ^2;\;\rw^2\equiv 1.
\end{equation}
If $\rw\equiv 0$, another simple calculation yields
\begin{equation}\label{RN}
A(r)=1-\frac{\Lambda r^2}{3}+\frac{1}{r^2}+\frac{c}{r};\;\rw\equiv 0
\end{equation}
where $c$ is an arbitrary constant.
If, for any $r>0$, ${\rw'}(r)=0$ and ${\rw}^2(r)=1$ or $0$, then by uniqueness of solutions of ordinary differential equations, the solution must be~(\ref{AdeSit}) or~(\ref{RN}).\hfill $\blacksquare$
\begin{list}{\textit Fact \arabic{fact}:}
{\usecounter{fact}
\addtocounter{fact}{2}
\findent}
\item\label{Einstsp}  There is another known explicit solution; namely Einstein space, when $\Lambda=3/4$, $A=1-r^2/2$, and  $\rw=\sqrt{A}$.
\end{list}
\begin{list}{\textit Fact \arabic{fact}:}
{\usecounter{fact}
\addtocounter{fact}{3}
\findent}
\item \label{family} Given any $\Lambda$ and $\lambda$ there is an interval $I_\lambda=[0,r_\lambda)$ in which there exists a solution of
equations (\ref{Aeq}) and (\ref{weq}) with the following properties:
\begin{list}{(\Roman{prop}):}
{\usecounter{prop}
\lindent}
\item $(A,\rw^2,\rw')\rightarrow (1,1,0)$ as $r\searrow 0$ and $\lim_{r\searrow 0}\rw''(r)=-\lambda$.
\item The solution is analytic in the interior of $I_\lambda$
and $\mathcal{C}^{2+\alpha}$ in $I_\lambda$ for a small $\alpha>0$.
\item The solutions depend continuously on $\lambda$.
\end{list}
\end{list}
A proof of \textit{Fact~\ref{family}} in the case $\Lambda = 0$ can be found in \cite{js91}.  The same proof is valid with minor modification in the general case $\Lambda\ne 0$.\\

We define the region
\[\Gamma=\{(r,A,\rw,{\rw'}):r>0,\;A>0,\; {\rw}^2\le 1\; \mathrm{and} \;(\rw,{\rw'})\not=(0,0)\}.\]
and set $r_c$ to the smallest value of $r$ that satisfies $A(r_c)=0$ if such an $r$ exists and set $r_c=\infty$ if no such $r$ exists.  This gives the following:
\begin{list}{\textit Fact \arabic{fact}:}
{\usecounter{fact}
\addtocounter{fact}{4}
\findent}
\item\label{permexit} Suppose for some $r_0\ge 0$, $(r_0,A(r_0),\rw(r_0),\rw'(r_0))\in\Gamma$ but $(r_e,A(r_e),\rw(r_e),\rw'(r_e))\notin \Gamma$ for some $r_e\in(r_0,r_c)$.  Then
$(r,A(r),\rw(r),\rw'(r))\notin \Gamma$ for all $r\in [r_e,r_c)$.
\end{list}
\textbf{Proof of \textit{Fact~\ref{permexit}}}:  At $r_e$ one of the following must hold:
\newcounter{whenexit}
\begin{list}{(\arabic{whenexit}):}
{\usecounter{whenexit}
\lindent}
\item\label{wea} $A(r_e)=0$,
\item\label{web} $(\rw(r_e),\rw'(r_e))=(0,0)$, or
\item\label{wec} ${\rw(r_e)}^2>1$.
\end{list}
We now examine each of these cases.\\ \\
\textit{Case~\ref{wea}}.  $r_e=r_c$ and there is nothing to prove.\\ \\
\textit{Case~\ref{web}}.  From \textit{Fact}~\ref{wconst}, $\rw\equiv 0$, contrary to the hypotheses.\\ \\
\textit{Case~\ref{wec}}.  We assume the contrary; namely, that there exists an $r_1>r_0$ such that $\rw(r_1)=\pm 1$ with sign agreeing with that of $\rw$ when it leaves $\Gamma$.  Because the equations are smooth in the region $[r_0,r_1]$, $\rw$ is also smooth in this region.  It follows that there exists a $\bar\rho\in (r_0,r_1)$ that satisfies ${\rw}^2(\bar
\rho)=1$ and ${\rw'}\rw(\bar \rho)\ge 0$.  If ${\rw'}(\bar \rho)=0$, then by \textit{Fact}~2, ${\rw}^2\equiv 1$.  We may therefore assume that ${\rw'}(\bar \rho)\ne 0$.  Thus, there exists also an $\bar r\in (\bar \rho,r)$ such that
${\rw'}(\bar r)=0$ and $\rw''\rw(r)<0$; (see Figure~1).  But equation~(\ref{weq}) shows that this is impossible.
\begin{center}
Figure~1.
\end{center}
\setlength{\unitlength}{0.7mm}
\begin{picture}(0,100)(-40,0)\label{pexit}
\put (10,15){\vector(0,1){80}}
\put (0,50) {\vector(1,0){100}}
\put (10,70) {\line(1,0){90}}
\put (3.5,45){$(0,0)$}
\put (101,49){$r$}
\put (8,95.5){w}
\put (5,68){$1$}
\qbezier (20,20) (40,130)(80,55)
\put (30,50){\dashbox(0,10)}
\put (27.1,45){$r_0$}
\put (34.3,50){\dashbox(0,20)}
\put (33.4,45){$\bar\rho$}
\put (73,50){\dashbox(0,17.1)}
\put (72,45){$r_1$}
\put (60,50){\dashbox(0,32)}
\put (58,45){$r_e$}
\put (80,50){\dashbox(0,5)}
\put (80,45){$r_c$}
\put (25,20){$\rw(\bar r)>1,\;\rw'(\bar r)=0$, $\rw''(\bar r)\le 0$}
\put (51,50){\dashbox(0,35.45)}
\put (49.5,45){$\bar r$}
\end{picture}
\par
\indent
To avoid repeating the same argument several times, we state the following basic calculus Lemma:
\begin{Lma}\label{ffinite}
Suppose $\lf_{r\searrow 0}(rf')(r)>0$.  Then $\lim_{r\searrow 0}f(r)=-\infty$. Similarly, if $\lp_{r\searrow 0}(rf')<0$, then $\lim_{r\searrow 0}f(r)=\infty$.
\end{Lma}

\section{Extending Solutions}
\label{extend}
\setcounter{equation}{0}
$\;$

Our first result is the following: 
\begin{thm}\label{Main1}
Let $\bar r$, $\bar A>0$ be arbitrary and let $A(r)$, $\rw(r)$ be any solution of (\ref{Aeq}) and 
(\ref{weq}) satisfying $A(\bar r)=\bar A$, valid 
in a neighborhood $(r_0,\;r_1)$ of $\bar r$.  Define $r_0\ge 0$ be the smallest $r$ to which the solution
can be extended.  
If $\lp_{r \searrow r_0} A(r) > 0$, then  $r_0=0$.
\end{thm}
\noindent
\textbf{Proof}:  It follows from standard theorems that $r_0>0$ only 
if one of the following holds:

\begin{list}{(\Alph{prop}):}
{\usecounter{prop}
\lindent}
\item\label{a-} $\lf_{r\searrow r_0} A(r) \le 0$,
\item\label{wblow}$\lp_{r\searrow r_0} \rw^2(r)=\infty$,
\item\label{wpblow}$\lp_{r\searrow r_0} {\rw'}^2(r)=\infty$, or

\item\label{ablow}$\lp_{r\searrow r_0} A(r)=+\infty$.
\end{list}
We eliminate all of these possibilities in each of the following cases:
\begin{list}{(\arabic{prop}):}
{\usecounter{prop}
\lindent}
\item\label{a>1}$\lp_{r\searrow r_0}A(r)>1$,
\item\label{awithlim}$\lim_{r\searrow r_0}A(r)$ exists and $0<\lim_{r\searrow r_0}A(r)\le 1$, and
\item\label{anolim}$\lf_{r\searrow r_c}A(r)<\lp_{r\searrow r_0}A(r)\le 1$.
\end{list}
\textit{Case~\ref{a>1}}.  In Lemma~\ref{0Agt1} we will prove that $\lim_{r\searrow r_0}A(r)=A_0$ exists and that $A_0>1$.  Thus, $r_0$ cannot be greater than $0$ on account of condition (A).  The other possibilities are eliminated according to the scheme shown in Figure~2 in the case $A_0=\infty$ and as shown in Figure~3 for the case $A_0<\infty$.  Figures~2 and~3 should be read as follows: at each node of the tree, we assume that everything up to and including the root is true.  What is cited in parentheses excludes the possibility that under the assumptions, $r_0>0$.  It is obvious that the terminal branches of these trees exhaust all of conditions (B), (C), and (D).\\ \\
\textit{Case~\ref{awithlim}}.  All of the possibilities are eliminated as shown in Figure~3.\\ \\
\textit{Case~\ref{anolim}}.  We assume $r_0>0$.  Then for any $M>0$, there exist $\rho_M$ that satisfy $0<r_0<\rho_M$, $0<A(\rho_M)\le 1$, and $\rho_M A'(\rho_M) > M$.  For $M$ sufficiently large, 
\[rA' + 2A{\rw'}^2 - 1 + A +\frac{(\rw^2-1)^2}{r^2} + \Lambda r^2|_{(r=\rho_M)} >0.\] 
This contradicts equation~(\ref{Aeq}) and establishes that $r_0=0$.  \hfill $\blacksquare$
$\;$\\ \\ \\
\begin{center}
Figure~2
\end{center}
\setlength{\unitlength}{1.0mm}
\begin{picture}(0,100)(25,0)\label{Ainffig}
\usecounter{figs}
\put (80,95){$\lim_{r\searrow r_0}A(r)=\infty$}
\put (110,90){\vector(1,-1){20}}
\put (90,90){\vector(-1,-1){20}}
\put (55,65){$\lp_{r\searrow r_0}{\rw'}^2(r)=\infty$}
\put (120,65){$\lp_{r\searrow r_0}{\rw'}^2(r)<\infty$}
\put (70,60){\vector(0,-1){20}}
\put (75,60){\vector(4,-3){27}}
\put (65,30){\vector(-1,-1){20}}
\put (75,30){\vector(1,-1){20}}
\put (55,35){$\lim_{r\searrow r_0}{\rw}(r)$ exists}
\put (105,35){$\lim_{r\searrow r_0}{\rw}(r)$ does not exist}
\put (120,30){(Lemma~\ref{0osc})}
\put (125,60){(Lemma~\ref{0Ainf})}
\put (30,5) {$\lim_{r\searrow r_0}\rw(r)<\infty$}
\put (80,5){$\lim_{r\searrow r_0}\rw(r)=\infty$}
\put (35,0){(Lemma \ref{0Awlim})}
\put (85,0){(Lemma~\ref{0winf})}
\end{picture}
\pagebreak

\begin{center}
Figure~3
\end{center}
\setlength{\unitlength}{1.0mm}
\begin{picture}(0,100)(25,0)\label{Afinfig}
\usecounter{figs}
\put (80,95){$0<\lim_{r\searrow r_0}A(r)<\infty$}
\put (110,90){\vector(1,-1){20}}
\put (90,90){\vector(-1,-1){20}}
\put (55,65){$\lp_{r\searrow r_0}{\rw}^2(r)<\infty$}
\put (120,65){$\lp_{r\searrow r_0}{\rw}^2(r)=\infty$}
\put (70,60){\vector(0,-1){20}}
\put (75,60){\vector(4,-3){27}}
\put (65,30){\vector(-1,-1){20}}
\put (75,30){\vector(1,-1){20}}
\put (55,35){$\lp_{r\searrow r_0}{\rw'}^2(r)=\infty$}
\put (105,35){$\lp_{r\searrow r_0}{\rw'}^2(r)<\infty$}
\put (97,30){(Equations smooth at $r_0$)}
\put (125,60){(Lemma~\ref{0winf})}
\put (30,5) {$\lim_{r\searrow r_0}\rw(r)$ exists}
\put (75,5){$\lim_{r\searrow r_0}\rw(r)$ does not exist}
\put (35,0){(Lemma \ref{0Awlim})}
\put (85,0){(Lemma \ref{0osc})}
\end{picture}
$\;$\\ \\ \\
It remains to prove Lemmas~\ref{0winf},~\ref{0osc},~\ref{0Awlim},~\ref{0Ainf}, and~\ref{0Agt1}.
\begin{Lma}\label{0winf}
Suppose $\lim_{r\searrow r_0} A(r)>0$ and $\lp_{r\searrow r_0} \rw(r)=\pm \infty$.
Then $r_0=0$.
\end{Lma}
\noindent  
\textbf{Proof}:   We assume that $r_0>0$ and will arrive at a contradiction.  Without loss of generality, we may also assume that $\lp_{r \searrow r_0} \rw(r) = + \infty$.  Then, from an argument similar to that used to prove \textit{Fact}~\ref{permexit}, we have $\lim_{r\searrow r_0}\rw(r)=+\infty$.  Consequently,  $\lf_
{r\searrow r_0} \rw'(r)=-\infty$.  

We assert that there exists a neighborhhod $U=(r_0,r_0+\epsilon)$ such that $\rw'(r)<0$ for all $r\in U$.  Indeed, equation~(\ref{weq}) implies $\rw''(\rho)>0$ for any $\rho$ that satisfies $\rw'(\rho)=0$ provided that $\epsilon$ is sufficiently small so that $\rw(1-\rw^2)<0$ for all $r\in U$ and that $\rho\in U$.  Consequently, $\rw'$ can have only one sign near $r_0$.  Clearly this sign must be negative.   

We next prove that 
\begin{equation}\label{locwp}
\lim_{r\searrow r_0}\rw'(r)=-\infty.
\end{equation}
To this end, we consider the following equation which is obtained easily from equations~(\ref{Aeq}) and (\ref{weq}):
\begin{equation}\label{veq}
r(A\rw')'+2{\rw'}^2(A\rw')+\frac{\rw(1-\rw^2)}{r}=0.
\end{equation}
It is clear from Equation~(\ref{veq}) that $(A\rw')'>0$ for all $r\in U$.   This 
implies that $\lim_{r\searrow r_0}A\rw'(r)$
exists.  Because $\lim_{r\searrow r_0}A(r)$ exists and is nonzero,
$\lim_{r\searrow r_0}\rw'(r)$ also exists.  Since  $\lim_{r\searrow r_0} \rw(r)=+\infty$,
the only possible limit for $\rw'$ is $-\infty$.  This establishes equation~(\ref{locwp}).

To complete the proof, we write equation~(\ref{weq}) in the following form:
\[\frac{r^2A\rw''}{\rw(\rw^2-1)} + \rw'r[\frac{1-
A-\Lambda r^2}{\rw(\rw^2-1)}- \frac{(\rw^2-1)}{\rw r^2}] - 1 = 0.\] 
For all $r\in U$, the
term inside the square brackets is negative.  Also, equation~(\ref{locwp}) gives $\eta>0$ such that 
$ \rw'(r) < -\eta <0$ for all $r\in U$.  It follows that $\rw'' <0$ throughout $U$.  But this contradicts equation~(\ref{locwp}).  \hfill $\blacksquare$ 
\begin{Lma}\label{0osc}
Suppose $\lf_{r\searrow r_0} A(r) > 0$ and $\lf_{r\searrow r_0} \rw(r) < \lp_{r\searrow r_0} \rw(r).$  Then $r_0 = 0$.
\end{Lma}
\noindent
\textbf{Proof}:  We assume $r_0 > 0$ and will obtain a contradiction.  As in \textit{Fact}~\ref{permexit}, equation~(\ref{weq}) implies $\lp_{r\searrow r_0}\rw^2(r)\le 1$.

We now claim that for any $\epsilon>0$ and $M>0$, there exist $\hat r(\epsilon,M)$ close to $r_0$ with $|\rw'(\hat
r)|<\epsilon$ and $|\rw''(\hat r)|>M$.  Because $\lim_{r\searrow r_0}\rw(r)$ does not exist, there exist sequences 
$\{r_n\}\searrow r_0$ such that $\rw'(r_n)=0$,
$\rw(r_n)\rightarrow \lp_{r\searrow r_0}\rw(r)$, and $\{s_n\}\searrow r_0$ such that $\rw'(r_n)=0$,
$\rw(s_n)\rightarrow \lf_{r\searrow r_0}\rw(r)$.  Without loss of generality, we take $r_n<s_n<r_{n-1}$.  The Mean Value Theorem gives $t_n\in (s_n,r_{n-1})$ such that
\[\rw'(t_n)=\frac{\rw(r_{n-1})-\rw(s_n)}{r_{n-1}-s_n}.\]
Clearly, this goes to $\infty$ as $n\rightarrow \infty$. 
Similarly, we can find $u_n$, $r_n<u_n<s_n$ such that $\rw'(u_n)\rightarrow -\infty$.  Now, for any $\epsilon>0$, we define $V_n(\epsilon)=(a_n,b_n)$ to be the largest open neighborhood of $r_n$ in which $|\rw'|<\epsilon$.  Each $V_n$ is nonempty.  Also, for sufficiently large $n$, $V_n$ contains neither $t_n$ nor $u_n$.  Thus $(b_n-a_n)\rightarrow 0$. 
Applying the Mean Value Theorem again shows that there exist $\hat r_n\in V_n$ that satisfy
\[\rw''(\hat r_n)=\frac{\rw'(r_n)-\rw'(a_n)}{(r_n-a_n)}=\frac{\epsilon}{(r_n-a_n)}\rightarrow \infty\;\mathrm{as}\;n\nearrow \infty.\]
This proves the claim.   

Finally, we write equation~(\ref{weq}) as
\begin{equation}\label{wbyA}  
[\bar r^2\rw'' 
+\frac{\bar r\rw'}{A}(1-\frac{(\rw^2-1)^2} {\bar r^2} - \Lambda
\bar r^2) - \bar
r\rw' +\frac{\rw(1-\rw^2)}{A}]_{r=\bar r(\epsilon,M)} = 0.
\end{equation}
Because $\rw$ and $\bar r$ are bounded, $\epsilon$ can be chosen to be sufficiently small and $M$ chosen to be sufficiently large so that the first term in equation~(\ref{wbyA}) dominates. Therefore the left side of
equation~(\ref{wbyA}) will be nonzero, giving a contradiction.  The result follows.  \hfill $\blacksquare$ 

\begin{Lma}\label{0Awlim}
Suppose $\lim _{r\searrow r_1} A(r) > 0$ exists.  Suppose, in addition, that $\lp_{r\searrow r_0} {\rw'}^2(r) = \infty$, and that
$\lim_{r\searrow r_0} \rw(r)=\rw_0$ is finite.
Then $r_0=0$.
\end{Lma}
\noindent
\textbf{Proof}: We assume $r_0>0$ and will obtain a contradiction.  
Clearly, 
\[\lp_{r\searrow r_0}\Phi(r)<1\]
and, without loss of generality, we assume that $\lp_{r\searrow r_0}\rw'(r)=+\infty$. Then, 
\[\lp_{r\searrow r_0}\ln(\rw')=+\infty.\]
This implies  that $\lf_{r\searrow r_0}(\ln(\rw))'=-\infty$;
i.e, for any $M>0$ there exists an $\bar r$ near $r_0$ with $\rw'(\bar r)>1$ and  $\rw''(\bar
r)/\rw'(\bar r)<-M$.  We now choose $M$ to be sufficiently large so that  
\begin{equation}\label{wbyw'}
[r^2 A \frac{\rw''}{\rw'}+r \Phi +\frac{\rw(1-\rw^2)}{\rw'}]_{r=\bar r}<0.
\end{equation}
This is possible because the first term on the left can be made large and  negative, the second term will be at most a bounded positive number, and the
third term will be bounded.  However, inequality~(\ref{wbyw'}) contradicts equation
(\ref{weq}).  The result follows.  \hfill $\blacksquare$ 
  
\begin{Lma}\label{0Ainf}
 Suppose $\lim_{r\searrow r_0} A(r)=+\infty$ and $\lp_{r\searrow r_0}
{\rw'}^2(r) < +\infty$.  Then $r_0=0$.
\end{Lma}
\noindent
\textbf{Proof}:  Equation~(\ref{Aeq}) can be written as 
\begin{equation}\label{AbyA}
\frac{rA'}{A}+1+2{\rw'}^2 = \frac{1}{A}-\frac{(\rw^2-1)^2}{r^2A} -\frac{\Lambda r^2}
{A}.
\end{equation}

We assume $r_0>0$ and will obtain a contradiction.  The right hand side of equation~(\ref{AbyA}) approaches $0$ whereas there exists some positive $M$ such that $1+2{\rw'}^2<r_0M$ in a neighborhood $U=(r_0,\;r_0+\epsilon)$.  
Thus, 
\begin{equation}\label{a'bound}
A'>-MA\;\;\;\mathrm{for\;all}\;r\in U.
\end{equation}
Integrating inequality~(\ref{a'bound}) on any interval $(r,r_2)\subset U$ gives 
\begin{equation}\label{aexp}
A(r)<A(r_2)e^{M(r_2-r)}\;\;\;\mathrm{for\; all}\; r\in U.
\end{equation}
Taking the limit in inequality~(\ref{aexp}) as $r \searrow r_0$ yields \[\lim_{r\searrow r_0}
A(r)<A(r_2)e^{M\epsilon}<\infty.\]
This  contradicts our hypothesis.  \hfill $\blacksquare$

\begin{Lma}\label{0Agt1}
Suppose there exists some $\rho \in (r_0,\bar r)$ such that $A(\rho)=\bar A\ge 1-\Lambda \rho^2/3$.    
Then $\lim_{r\searrow r_0} A(r)$ exists and $\lim_{r\searrow r_0} A(r) \ge 1 - 
\Lambda r_0^2/3$.
\end{Lma}
\noindent
\textbf{Proof}:   We define
\begin{equation}\label{mu}
\mu=r(1-A-\Lambda r^2/3).
\end{equation}
A simple calculation using equation~(\ref{Aeq}) yields
\begin{equation}\label{mu'} 
\mu'=1-A-rA'-\Lambda r^2=\frac{(\rw^2-1)^2}{r^2}+2A{\rw'}^2 \ge 0.
\end{equation}
If $(\rw^2(\rho),\rw'(\rho))=(1,0)$, then
\begin{equation}\label{Abar}
A=1 - \frac {\Lambda r^2}{3} + \frac{\rho}{r}[\bar A + \frac{\Lambda  \rho
^2}{3} - 1],\;\; \;{\rw}^2\equiv  1
\end{equation} 
is the unique solution of equations~(\ref{Aeq}) and (\ref{weq}) that satisfies $\rw(\rho)=1$, $\rw'(\rho)=0$, 
$A(\rho)=\bar A$.  However, our assumptions imply that the term in the square brackets of equation~(\ref{Abar}) is
nonnegative.  This term is also obviously constant with respect to $r$.  Therefore, the solution equation~(\ref{Abar}) is valid down to $r_0=0$ and
$\lim_{r\searrow 0}A(r)$ exists and equals either $1$ or $\infty$. 

To finish the proof, we need to consider the remaining
case; namely, the case when 
\[\rw(1-\rw^2)(\rho)\ne 0.\]  We first note that this implies $\mu'(\rho)/\rho> 0$.  Also, our assumptions imply that  
$\mu(\rho)/\rho^2\le 0$.  Therefore 
\[(\mu/r)'(\rho)=\mu'(\rho)/\rho-\mu(\rho)/\rho^2> 0.\]
We now suppose that there
exist $\hat r \in (r_0,\rho)$ such that  $(\mu/r)'(\hat r)=0$. 
$\hat r$ can always be chosen so that $\mu(\hat r)/\hat r<0$.  Then, 
\begin{equation}\label{mucontr}
\mu'(\hat r)=(r \mu/r)'(\hat r)=\hat r(\mu/r)'(\hat r)+(\mu(\hat r)/\hat r)<0.
\end{equation}
Equation~(\ref{mucontr}) contradicts equation~(\ref{mu'}).  It follows that in the interval $(r_0,\rho)$, $(\mu'/r)'>0$; i.e.,
\begin{equation}\label{A'int}
A'=-2 \Lambda r/3 -(\mu/r)'<-2\Lambda r/3.
\end{equation}
Since $A'$ is bounded from above, it follows that $\lim_{r\searrow r_c}A(r)$ exists.  Also from equation~(\ref{A'int}) and the fact that $A(\rho)\ge 1-\Lambda\rho^2/3$ it is clear that $\lim_{r\searrow r_c}A(r)\ge 1-\Lambda r_0^2/3$.  \hfill $\blacksquare$

\section{Behavior at the Origin}
\subsection{A is less than 1}
\label{A<1}
\setcounter{equation}{0}
$\;$

We prove the following:
\begin{thm}\label{Mainpart}
Suppose $(A(r),\;\rw(r))$ is a solution of equations~(\ref{Aeq}) and (\ref{weq}) valid down to $r=0$ and that there exist $A_0$ and $A_1$ such that 
\[0<A_0<\lf_{r\searrow 0}A(r)\le \lp_{r\searrow 0}A(r)=A_1\le 1.\]
Then 
\[\lim_{r\searrow 0}(A(r),\rw^2(r),\rw'(r))=(1,1,0).\]
\end{thm}
\noindent
Throughout this section, we assume $(A(r)$, $\rw(r))$ is a solution that satisfies the hypotheses of Theorem~\ref{Mainpart}.
We will first prove that $\lim_{r\searrow 0}{\rw}^2(r)=1$ and then use this together with Lemma~\ref{epsilon} to prove that $\lim_{r\searrow 0}\rw'(r)$ exists.  Theorem~\ref{Mainpart} will follow.\\ 

We begin with the following
\begin{Lma} $\lim_{r\searrow 0}\rw^2(r)=1$.
\end{Lma}
\noindent
\textbf{Proof}:  Equation~(\ref{weq}) shows, by arguments similar to those used to prove \textit{Fact}~\ref{permexit} or from \textit{Fact}~\ref{wconst}, that $\lim_{r\searrow 0}\rw(r)$ exists whenever there exists an $\bar r$ such that $\rw^2(\bar r)\ge 1$.  Consequently, we may assume that there exists a $\rho>0$ such that $\rw^2(r)<1$ whenever $r\in(0,\rho)$.

Also, it is easy to see from equation~(\ref{Aeq}) and Lemma~\ref{ffinite} that the assumption that $\lf_{r\searrow 0}A(r)<1$ forces
\[\lp_{r\searrow 0}\rw^2(r)=1.\]
  We now assume, without loss of generality, that 
\[-1\le\lf_{r\searrow 0}\rw(r)<\lp_{r\searrow 0}\rw(r)\le 1.\]
Equation~(\ref{weq}) implies that either $\lf_{r\searrow 0}\rw(r)\le 0$
or $\lp_{r\searrow 0}\rw(r)\ge 0$ since $\rw''\rw(r)<0$ whenever ${\rw'}(r)=0$.  There exists, therefore, a sequence $\{r_n\}\searrow 0$ such
that $\rw(r_n)\rightarrow \lp_{r\searrow 0}\rw(r)$ and ${\rw'}(r_n)=0$.  There also exists a sequence
$\{s_n\}$ such that
\begin{equation}
(A{\rw'})(s_n)\searrow -\infty\;\mathrm{and}\;(A{\rw'})'(s_n)=0.
\end{equation}
We choose any
$\delta\in(0,1)$, $\tilde A_1>A_1$, and define
\begin{equation}\label{cdef}
c=\max\{\sqrt{\tilde A_1/(3\delta A_0)},\sqrt{1-A_0},1/\sqrt{2A_0}\}.
\end{equation}
Also, for each $n$, we define 
\[r^0_n=\min\{r>r_n:\rw(r)=0\}\]
and
\[r^\delta_n=\min\{r>r_n:\rw(r)=\delta\}.\]
If $\rw(r_n)>1-cr_n$, we define 
\[t_n=\min\{r>r_n:\rw(r)=1-cr\}\]
whereas if $\rw(r_n)\le 1-cr_n$, we set $t_n=r_n$  (See Figure 3.)
\begin{center}
Figure~3.
\end{center}
\setlength{\unitlength}{0.7mm}
\begin{picture}(0,100)(-15,0)\label{oscpic}
\put (10,0){\vector(0,1){95}}
\put (0,50) {\vector(1,0){100}}
\put (10,70) {\line(1,0){90}}
\put (3.5,45){$(0,0)$}
\put (101,49){$r$}
\put (8,95.5){w}
\put (5,68){$\delta$}
\put (10,90){\line(1,0){90}}
\put (5,88){$1$}
\put (10,90){\line(4,-1){95}}
\put (95,63){$\rw=1-cr$}
\qbezier (20,20) (40,130)(60,60)
\put (44.5,50){\dashbox(0,37)}
\put (43.5,45){$r_n$}
\put (53,50){\dashbox(0,29)}
\put (52,45){$t_n$}
\put (56.76,50){\dashbox(0,20)}
\put (55.2,51.5){$r_n^\delta$}
\qbezier (60,60)(80,0)(80,80)
\put (62.5,45){$r_n^0$}
\put (63.5,49){\dashbox(0,2)}
\put (22,30){$\rw(r)$}
\end{picture}
$\;$\\
We will prove that, for sufficiently large $n$, there can be no $s_n$.  This will be our contradiction.

From equation~(\ref{veq}) it is clear that for each $n$, $s_n\in[r_m,r^0_m]$ for some $m$.  No generality is lost by assuming
$n=m$. Also, it is obvious that for sufficiently large $n$, $t_n<r^\delta_n$.  We now consider the
three intervals in which $s_n$ could possibly lie:

\begin{list}{(\arabic{prop}):}
{\usecounter{prop}
\lindent}
\item\label{high}$s_n\in[r_n,t_n]$,
\item\label{middle}$s_n\in(t_n,r^\delta_n)$, and
\item\label{low}$s_n\in[r^\delta_n,r^0_n]$.
\end{list}
\textit{Case~\ref{high}}.  In this interval $\rw(1-{\rw}^2)/r=\rw(1+\rw)(1-\rw)/r<2c$; i.e.,
$\rw(1-{\rw}^2)/r$ is bounded from above.  Since we
are assuming ${\rw'}(s_n)\searrow -\infty$, equation~(\ref{veq}) implies $s_n$ cannot be in this interval.\\ \\
\textit{Case~\ref{middle}}.  If $t_n=r_n$ we can ignore this case.  Otherwise, we note that from the definition of $t_n$, it follows that
\begin{equation}\label{w'tn}
{\rw'}(t_n)\le -c.
\end{equation}
Also, throughout the interval $(t_n,r^\delta_n)$,
\[\Phi=1-A-(1-{\rw}^2)^2/r^2-\Lambda r^2<1-A_0-c^2<0.\]
Substituting this into equation~(\ref{weq}) gives 
\begin{equation}\label{w''<0}
r\rw''(r)<-\frac{\rw(1-{\rw}^2)}{rA}= -\rw(1+\rw)\frac{1-\rw}{rA}<-\frac{\delta c}{\tilde A_1}.
\end{equation}
We now consider the function 
\begin{equation}\label{q}
q(r)=2rA_0{\rw'}^3+\rw(1-{\rw}^2).
\end{equation}
\begin{equation}\label{q'}
q'(r)={\rw'}(2A_0{\rw'}^2+6A_0r{\rw'}\rw''+1-3{\rw}^2).
\end{equation}
Since $\rw''<0$, we have
\begin{equation}\label{w'attn}{\rw'}(r)<{\rw'}(t_n).
\end{equation}
Substituting equations~(\ref{w'tn}), (\ref{w''<0}) and (\ref{w'attn}) into equation~(\ref{q'}) yields 
\begin{equation}\label{q'est}
q'(r)<{\rw'}(r)[6A_0r\rw''(r){\rw'}(t_n)-2]\le {\rw'}(r)[6A_0\delta c^2/\tilde A_1-2]<0.
\end{equation}
The last inequality follows from equation~(\ref{cdef}).  Equations~(\ref{veq}), (\ref{q}), (\ref{q'est}), (\ref{w'tn}), and (\ref{cdef}) now yield
\begin{eqnarray}\label{qest}
r^2(A{\rw'})' &=&-2rA{\rw'}^3-\rw(1-\rw^2)\nonumber\\
 & >&-2rA_0{\rw'}^3-\rw(1-\rw^2)\nonumber\\
 &=&-q(r)>-q(t_n)\nonumber\\
 & >&2A_0rc^3-rc\nonumber\\
 &=& rc(2A_0c^2-1)>0;
\end{eqnarray}
i.e., $s_n$ cannot be in the interval $(t_n,r_n^\delta)$.\\ \\
\textit{Case~\ref{low}}.  For $n$ sufficiently large, equation~(\ref{weq}) gives
\begin{eqnarray}\label{nosn}
{\rw'}(r)&=&{\rw'}(t_n)+\int_{t_n}^r \rw''(\rho) \rd \rho\nonumber\\
&=&{\rw'}(t_n)+\int_{t_n}^r(\frac{-\rw(1-{\rw}^2)}
{\rho^2 A}-\frac{\Phi {\rw'}}{\rho A}) \rd \rho\nonumber\\
 &\le&\int_{t_n}^r-\frac{\Phi {\rw'}}{\rho A} \rd \rho\nonumber\\
 &\le&\frac{\epsilon}{r A_1}\int_{t_n}^r{\rw'} \rd
\rho\nonumber\\
 &=&\frac{\epsilon}{r A_1}[\rw(r)-\rw(t_n)]
\end{eqnarray}
for any $r\in[r_n^\delta,r_n^0]$.  The last inequality follows for arbitrary positive  $\epsilon<c^2-(1-A_0)$ from equation~(\ref{cdef}).  This is because, for any such $\epsilon$,
\[\Phi=1-A-(1-{\rw}^2)^2/r^2-\Lambda r^2<1-A_0-c^2<-\epsilon.\]
throughout the interval $[t_n,r_n^0]$ for sufficiently large $n$.  We have also used the fact that $\rw'< 0$ in this same interval.

We now choose an arbitrary $\tilde\delta\in(\delta,1)$.  Because $\lim_{n\nearrow \infty}\rw(t_n)=1$, for sufficiently large $n$ and arbitrary $r\in[r_n^\delta,r_n^0]$,
\begin{equation}\label{winterval}
\rw(r)-\rw(t_n)<-(1-\tilde\delta).
\end{equation} 
Substituting equation~(\ref{winterval}) into inequality~(\ref{nosn}) yields
\begin{equation}\label{w'comp}
{\rw'}(r)<-\frac{\epsilon(1-\tilde\delta)}{rA_1}.
\end{equation}
Finally, we substitute inequality~(\ref{w'comp}) into equation~(\ref{veq}) to get,
for large $n$,
\begin{equation}\label{nosncomp}
r(A{\rw'})'(r)=-2A{\rw'}^3(r)-\frac{\rw(1-{\rw}^2)}{r}>\frac{2A(r)\epsilon^3(1-\tilde\delta)^3}{r^3A_1^3}-\frac{1}{r}.
\end{equation}
It is clear that for sufficiently large $n$, the first term on the right side of inequality~(\ref{nosncomp}) dominates.  Thus, for sufficiently large $n$,
\[(A{\rw'})'>0\;\;\;\mathrm{for\;all}\;r\in[r_n^\delta,r_0^\delta];\]
i.e., $s_n$ is not in the interval $[r_n^\delta,r_n^0]$.  \hfill $\blacksquare$ \\

Having established the existence of $\lim_{r\searrow 0}\rw(r)$ next establish the existence of
$\lim_{r\searrow 0}{\rw'}(r)$.  Without loss of generality (\textit{Fact}~\ref{reflect}) we may assume that $\lim_{r\searrow 0}\rw(r)=1$.  From equation~(\ref{weq}), the fact that $\rw(0)=1$, and \textit{Fact} \ref{wconst} it follows that $\rw'(r)\ne 0$ in a neighborhood of $0$ unless $\rw\equiv 1$.  If $\rw\equiv 1$, then obviously $\lim_{r\searrow 0}\rw'(r)=0.$  We are left with only two other possibilities:

\begin{list}{(\arabic{prop}):}
{\usecounter{prop}
\lindent}
\item\label{w<1}For any $\epsilon>0$, there exists an $r_0>0$ with $1-\epsilon<\rw(r)<1$ and $\rw'(r)<0$ for all $r\in(0,r_0)$, and
\item\label{w>1}For any $\epsilon>0$, there exists an $r_0>0$ with $1<\rw(r)<1+\epsilon$ and $\rw'(r)>0$ for all $r\in(0,r_0)$.
\end{list}
We consider only Case~\ref{w<1}; the proof for Case~\ref{w>1} being similar.
The existence of $\lim_{r\searrow 0}\rw'(r)$ is a consequence of the following:
\begin{Lma}\label{epsilon}
There exists an $r_0$ independent of $\epsilon$ such that if for any $\epsilon>0$  and for any $b\in(0,r_0)$, 
\[1-\epsilon b\le \rw(b)<1,\]
then 
\[1-\epsilon r\le\rw(r)<1\]
for all $r\in(0,b)$.
\end{Lma}
\noindent
\textbf{Proof}:  For any $\epsilon>0$ we define 
\[U_\epsilon=\{r\in [0,b]:\rw(s)\ge 1-\epsilon s\;\mathrm{for\;all}\;s\in[0,r]\}\]
 and define also 
\[a_\epsilon=\sup \{r\in U_\epsilon\}.\]
$U_\epsilon$ is nonempty since it contains $0$..  It is also clear that $U_\epsilon$ is closed; i.e., $a_\epsilon\in U_\epsilon$.  Next, in the interval
$[a_\epsilon,b]$, we define
\begin{equation}\label{g}
g(\epsilon,r)=1-\epsilon r-\rw.
\end{equation}
Since $\epsilon$ is constant, we denote $g(\epsilon,r)$ also by $g(r)$.  We now prove $a_\epsilon=b$.  For otherwise, there exist $\tilde c\in(a_\epsilon,b)$ such
that $g(\tilde c)>0$.  Let $c\in[0,b]$ be where $g$ assumes its maximum.  Since $g(a)=g(b)=0$ and $g(\tilde c)>0$, $c\in (0,b)$ and, consequently, $g'(c)=0$ and $g''(c)\le 0$;
i.e., $\rw(c)<1-\epsilon c$, ${\rw'}(c)=-\epsilon$ and $\rw''(c)\ge 0$.  Now, 
equation~(\ref{weq}) gives
\begin{eqnarray}\label{unifeps}
0&=&[r^2A\rw'' + r\Phi {\rw'} +\rw(1-{\rw}^2)]_{r=c}\nonumber\\
 &\ge &[c{\rw'}\Phi+\rw(1-{\rw}^2)]_{r=c}\nonumber\\
 &\ge & [-c\epsilon +\rw(1-{\rw}^2)]_{r=c}\nonumber\\
 &=& c[-\epsilon +\rw(1+\rw)\frac{(1-\rw)}{c}]_{r=c}\nonumber\\
 &>&c\epsilon[-1+\rw(1+\rw)]_{r=c}>0
\end{eqnarray}
provided $r_0$ is small enough so that $\rw(1+\rw)>1$ for all $r\in(0,r_0)$.  We have also used the fact that $\Phi<1$. From Equation~(\ref{unifeps}), it is obvious that there can be no $c$.  The result follows.  \hfill $\blacksquare$
\begin{Lma}\label{wp}
$\lim_{r\searrow 0}\rw'(r)$ exists and is finite.
\end{Lma}
\noindent
\textbf{Proof}:  We refer to equation~(\ref{g}) which, for each solution,  is defined on $[0,\infty]\times (0,r_c)$ for some $r_c>0$.  We now define the set
\[O=\{\epsilon\ge 0:\;\mathrm{there}\; \mathrm{exist}\; \rho_\epsilon>0\; \mathrm{such}\;\mathrm{that}\;
g(\epsilon,r)>0\;\mathrm{for}\;\mathrm{all}\; r\in(0,\rho_\epsilon)\}.\]
In addition, we define
\begin{equation}\label{epsdef}
\bar{\epsilon}=\sup\{\epsilon\in O\}.
\end{equation}
If there exist $\epsilon$ and $\rho_\epsilon$ such that $g(\epsilon,r)\equiv 0$ in $(0,\rho_0)$ then there is nothing to prove.  Consequently,  we assume this is not the case.

We first prove that $\bar\epsilon$ is well defined.  $O$ is nonempty since $0\in O$. Also, for any $\epsilon\in O$, if $\epsilon>1$, for all $r\in(0,\rho_\epsilon)$,
\[\frac{(1-\rw^2)^2}{r^2}>1\]
and, as a consequence of equation~(\ref{Aeq}), there exist $\eta>0$ such that in the same interval,
\[rA'<-\eta.\]
Lemma~\ref{ffinite} then implies $\lim_{r\searrow 0}A(r)=\infty$, contrary to our hypothesis.  We conclude that $\bar\epsilon\le 1$.  In particular, $\bar\epsilon<\infty$ and so is well defined.

 We claim also that $O$ is closed.  To prove this,  we choose arbitrary $\epsilon_0\notin O$.  There exists a $\rho_0\in(0,r_0)$ such that $g(\epsilon_0,\rho_0)\le 0$.  By Lemma~\ref{epsilon}, $g(\epsilon_0,r)\le 0$ for all $r\in (0,\rho_0)$. Because $\rw\not\equiv 1+\epsilon_0 r$ in a neighborhood of the origin, there exist $\tilde r\in(0,\rho_0)$ such that $g(\epsilon_0,\tilde r)<0$.  Now, for $\epsilon$ sufficiently close to $\epsilon_0$, $g(\epsilon,\tilde r)<0$ also.  Applying Lemma~\ref{epsilon} again gives $g(\epsilon,r)\le 0$ for all $r\in (0,\tilde r)$; i.e., $\epsilon\not\in O$ for $\epsilon$ sufficiently close to $\epsilon_0$.  This proves $O$ is closed.  As a consequence, $\bar\epsilon\in O$.

We now consider the two following possibilities:
\begin{list}{(\ref{w<1}\alph{prop}):}
{\usecounter{prop}
\lindent}
\item\label{e>0}$\bar\epsilon>0,$ and
\item\label{e=0}$\bar\epsilon=0$.
\end{list}
\textit{Case~\ref{w<1}a}.  Equation~(\ref{weq}) gives
\begin{eqnarray}\label{wbareps}
rA{\rw}'' &=&-\frac{\rw(1-\rw^2)}{r}-\rw'\Phi\nonumber\\
 &\le&-\rw(1+\rw)\frac{1-\rw}{r}-\rw'
\end{eqnarray}
whenever $\rw'<0$ since $\Phi<1$.  We prove that
\begin{equation}\label{lpeps}
\lp_{r\searrow 0}\rw'(r)\ge -\bar\epsilon.
\end{equation}
From equation~(\ref{lpeps}) it follows that if $\lf_{r\searrow 0}\rw'(r)<-\bar\epsilon$, for any $\eta>0$, we can find a sequence $\{r_n\}\searrow 0$ such that $\rw'(r_n)>-\bar\epsilon-\eta$ and $\rw''(r_n)= 0$.  We choose $\eta<\bar\epsilon/3$.  Now, on any such sequence,
\[\rw(1+\rw)\frac{1-\rw}{r}>\frac{3}{2}(\bar\epsilon+\eta)\]
 because $\rw(r_n)\nearrow 1$.  Equation~(\ref{wbareps}) then gives $\rw''(r_n)<0$.  Thus, we can find no such sequence and, as a consequence,  $\lim_{r\searrow 0}\rw'(r)$ exists.   Clearly $0\ge\lim_{r\searrow 0}\rw'(r)\ge-\bar\epsilon$ gives $\rw'$ a finite limit.

It remains to establish equation~(\ref{lpeps}).  To this end,  we define $\delta(r)$ by
\begin{equation}\label{deltadef}
\rw(r)=1-\bar\epsilon r-\delta(r).
\end{equation}
We first note that 
\begin{equation}\label{delta+}
\delta(r)\ge 0\;\mathrm{for\; all}\;r\in(0,r_0).
\end{equation}
Indeed, if not, then there exist $\tilde r\in(0,r_0)$ such that $\delta(r)<0$.  Substituting equation~(\ref{deltadef}) into equation~(\ref{g}) yields
\[g(\bar\epsilon,\tilde r)=-\delta (\tilde r)<0.\]
From Lemma~\ref{epsilon} it follows that $g(\bar\epsilon,r)\le 0$ for all $r\in (0,\tilde r)$; i.e., $\bar\epsilon\not\in O$, contradicting the fact that $\bar\epsilon\in O$.

We now claim that
\begin{equation}\label{lfdelta}
\lf_{r\searrow 0}\frac{\delta(r)}{r}=0.
\end{equation}
Equation~(\ref{delta+}) gives $\lf_{r\searrow 0}\delta(r)/r\ge 0$.  Now we assume that
\[\lf_{r\searrow 0}\delta(r)/r\ge 3\tilde\eta>0\]
where $\tilde\eta$ is arbitrarily small.  Then, for any $r$ sufficiently close to $0$, $\delta(r)>2\tilde\eta r$ and, therefore, 
\[g(\bar\epsilon+\tilde\eta,r)=-\tilde\eta  r+\delta(r)>\tilde\eta r>0;\]
i.e., $\bar\epsilon+\tilde\eta\in O$, contradicting the definition of $\bar\epsilon$.  It follows that $\lf_{r\searrow 0}\delta(r)/r\le 0$.  From equation~(\ref{delta+}) we conclude equation~(\ref{lfdelta}).

Finally, we choose a sequence $\{r_n\}\searrow 0$ such that $\delta(r_n)/r_n\searrow 0$.  The Mean Value Theorem yields a sequence $\{s_n\}\rightarrow 0$ such that
\[\rw'(s_n)=\frac{\rw(r_n)-1}{r_n}=\frac{-\bar\epsilon r_n-\delta(r_n)}{r_n}=-\bar\epsilon-\frac{\delta(r_n)}{r_n}.\]
Clearly, $\rw'(s_n)\rightarrow -\bar\epsilon$.  This establishes equation~(\ref{lpeps}) and completes the proof of Lemma~\ref{wp} in Case~\ref{w<1}a.\\ \\
\textit{Case~\ref{w<1}b}.  For any $\epsilon>0$, there exists a sequence $\{r^\epsilon_n\}\searrow 0$ such that for each $n$, $1-\epsilon r^\epsilon_n<\rw(r^\epsilon_n)<1$.  From the Mean Value Theorem and the fact that $A<A_1$, it follows that for any $\epsilon>0$, there exists a sequence
$\{r_n\}\searrow 0$ such that $-\epsilon/A_1<\rw'(r_n)<0$.  Therefore
\begin{equation}\label{aw'seq}
-\epsilon<A\rw'(r_n)<0.
\end{equation}
 Also, Lemma~\ref{epsilon} provides a $\rho>0$ such that
$1-\epsilon r<\rw(r)<1$ for all $r\in (0,\rho)$.  Equation~(\ref{veq}) then implies
\begin{equation}\label{aw'est}
(A\rw')'(r)>0\;\mathrm{for\; all}\;r\in(0,\rho)\;\mathrm{whenever}\;\rw'(r)<-3(\epsilon/A_0)^{1/3}.
\end{equation}
From equations~(\ref{aw'seq}) and (\ref{aw'est}) it follows that 
\[-\max\{\epsilon,3(\epsilon/A_0)^{1/3}\}<\rw'(r)<0\]
 whenever $r\in(0,\rho)$.  Since $\epsilon>0$ is arbitrary, the result follows.  \hfill $\blacksquare$\\

The next simple Lemma eliminates any ambiguity in defining ${\rw'}(0)$.  It does not depend
on our particular equations.
\begin{Lma}\label{w'0}
Whenever $\lim_{r\searrow 0}{\rw'}(r)$ exists and is finite and $\rw$ is differentiable at $r=0$, ${\rw'}$ is continuous at
$0$.
\end{Lma}
\noindent
\textbf{Proof}:  Our assumptions imply $\lim_{r\searrow 0}\rw(r)$ exists and is finite.  From the definition of a derivative, for any $\epsilon>0$ there exists a sequence $\{r_n\}\searrow
0$ such that
\begin{equation}\label{wp0eq}
|\frac{\rw(r_n)-\rw(0)}{r_n}-\rw'(0)|<\epsilon.
\end{equation}
The Mean Value Theorem yields a sequence $\{s_n\}$, $0<s_n<r_n$ such that
\begin{equation}\label{MVT}
{\rw'}(s_n)=\frac{\rw(r_n)-\rw(0)}{r_n}.
\end{equation}
Substituting equation~(\ref{MVT}) into equation~(\ref{wp0eq}) yields 
\[|{\rw'}(s_n)-\rw'(0)|<\epsilon;\]
i.e., $\lim_{n\nearrow \infty}{\rw'}(s_n)=\rw'(0)$.  Since we assume $\lim_{r\searrow 0}{\rw'}(r)$ to exist, it must also equal $\rw'(0)$.  \hfill $\blacksquare$ \\

We denote $\lim_{r\searrow 0}\rw'(r)$ by $\rw'_0.$\\ \\
\textbf{Proof of Theorem~\ref{Mainpart}}:  What remains is to proved that
$\lim_{r\searrow 0}A(r)=1$ and $\rw'_0=0$.  We first prove that $\lim_{r\searrow 0}A(r)$ exists.

Equation~(\ref{Aeq}) gives, for any sequence $\{r_n\}\searrow 0$ that satisfies $A'(r_n)=0$,
\begin{equation}\label{A'0seq}
A(r_n)=\frac{1-\frac{(1-\rw^2(r_n))^2}{r_n^2}-\Lambda r_n^2}{1+2{\rw'}^2(r_n)}.
\end{equation}
L'H\^{o}pital's rule applied twice to $(1-\rw^2)^2/r^2$ proves that the expression on the right side of equation~(\ref{A'0seq}) approaches
\[\frac{1-4{\rw'_0}^2}{1+2{\rw'_0}^2}\] 
as $n\rightarrow \infty$.  It follows that $\lim_{r\searrow 0}A(r)$ exists.  We denote it by $A_0$.
 
Now, equation~(\ref{Aeq}) gives
\[\lim_{r\searrow 0}(rA')=1-A_0-4{\rw'}_0^2-2A_0{\rw'}_0^2;\]
i.e., this limit exists.  We have, from Lemma~\ref{ffinite},
\begin{equation}\label{ra'}
\lim_{r\searrow 0}(rA')(r)=0
\end{equation}
and since ${\rw'_0}$ is finite,
\begin{equation}\label{raw'}
\lim_{r\searrow 0}(rA'{\rw'}^2_0)(r)=0.
\end{equation}
Because $\Phi_0=\lim_{r\searrow 0}\Phi(r)<1$, 
equation~(\ref{weq}) yields
\begin{equation}\label{raw'ne0}
\lim_{r\searrow 0}(rA\rw'\rw'')(r)=-\Phi_0{\rw'}^2_0+2{\rw'}^2_0>0
\end{equation}
unless $\rw'_0=0$.  Equations~(\ref{raw'ne0}) and~(\ref{raw'}) imply that
\begin{equation}\label{aw'deriv}
\rw'_0\lim_{r\searrow 0}(rA\rw')'(r)>0
\end{equation}
whenever $\rw'_0\ne 0$.  Lemma~\ref{ffinite} now gives $\rw'_0=0$.  Substituting this into equation~(\ref{Aeq}) and making use of equation~(\ref{ra'}) proves $A_0=1$.  \hfill $\blacksquare$
\subsection{A is greater than 1}
\label{A>>1}
$\;$

In this section, we prove the following result:
\begin{thm}\label{A>1}
Whenever there exists an $r_0$ such that $A(r_0)>1-\Lambda r_0^2/3$, then
$\lim_{r\searrow 0}A(r)=\infty$ and $\lim_{r\searrow 0}\rw'(r)=0$.
\end{thm}
\noindent
\textbf{Proof}:  The first part is easy to prove.  We consider equation~(\ref{mu}) which defined the function $\mu$  
\[\mu(r)=r(1-A-\frac{\Lambda r^2}{3}).\]
Now, $\mu'(r)=2A{\rw'}^2+\frac{(1-\rw^2)}{r^2}\ge 0$; i.e., $\mu$ is increasing in the interval $(0,r_0)$.  Thus $\lim_{r\searrow 0}\mu(r)$ exists.  Since
$\mu(r_0)<0$ by assumption, $\lim_{r\searrow 0}\mu(r)<0$; i.e., $\lim_{r\searrow 0}rA>0$.  This can only hold if
$\lim_{r\searrow 0}A(r)=\infty$.\\

To prove the second statement, let us assume for the moment that \\
$\lim_{r\searrow 0}\rw'(r)=\rw'_0$ exists.  There are two cases 
to consider:
\begin{list}{(\arabic{prop}):}
{\usecounter{prop}
\lindent}
\item\label{w=1}$\lim_{r\searrow 0}\rw^2(r)=1$ and
\item\label{wne1}$\lim_{r\searrow 0}\rw^2(r) \ne 1$.
\end{list}
\textit{Case~\ref{w=1}}.  We set $\rw_0=1$ and apply L'H\^{o}pital's rule to 
$(\rw^2-1)/r$.  The result is $\lim_{r\searrow 0}(\rw^2(r)-1)/r=2\rw'_0$.  If $\rw'_0=\infty$, then there exists an 
$\tilde r$ arbitrarily close to $0$ such that $\rw(\tilde r)>1$, $\rw'(\tilde r)>0$ and $\rw''(\tilde r)<0$. 
This contradicts equation~(\ref{weq}).  Similarly, $\rw'_0 \ne -\infty$.  We thus have $(\rw^2-1)/r$
bounded near $0$ and $(\rw'_0)^2<\infty$.

If $\rw'_0\ne 0$, then \textit{Fact}~\ref{wconst} and equation~(\ref{weq}) imply (as in \textit{Fact}~\ref{permexit}) that $\rw'$ has only one sign near $r=0$.  Without loss of generality, we assume $\rw'>0$.  Since $A(r) \rightarrow \infty$ as $r \rightarrow 0$, for $r$ 
near $0$, equation~(\ref{weq}) also gives, for small $r$,
\[rA\rw''(r)=A\rw'(r)+(\frac{(\rw^2-1)^2}{r^2}-1+\Lambda r^2)\rw'+\frac{\rw(\rw^2-1)}{r}>
\frac{1}{2}A\rw'(r)>0.\]
From this and Lemma~\ref{ffinite} it follows that  $\lim_{r\searrow 0}\rw'(r)=0.$\\ \\
\textit{Case~\ref{wne1}}.  We suppose $\lim_{r\searrow 0}{\rw'}^2(r)>2\epsilon>0$ where $\epsilon<1$ and will arrive at a contradiction.  For $r$ near
$0$, ${\rw'}^2(r)>\epsilon$ and $|\lim_{r\searrow 0}A\rw'(r)|=\infty$.  Multiplying 
equation~(\ref{veq}) by $A\rw'$
gives 
\begin{eqnarray}\label{vvpeq}
(A\rw')(A\rw')' &= &\frac{-2r{\rw'}^2(A\rw')^2-\rw(1-\rw^2)(A\rw')}{r^2}\nonumber \\
 & \le & \frac{-2\epsilon r(A\rw')^2-\rw(1-\rw^2)(A\rw')}{r^2}.
\end{eqnarray}
Now, equation~(\ref{Aeq}) implies $\lim_{r\searrow 0}r(rA)'=-\infty$. Lemma~\ref{ffinite} then gives 
\[\lim_{r\searrow 0}(rA)=\infty.\]
Therefore, for $r$ near $0$,
\begin{eqnarray}
-2\epsilon r(A\rw')^2-\rw(1-\rw^2)A\rw' &=&-\epsilon r(A\rw')^2-\epsilon r(A\rw')^2-\rw(1-\rw^2)A\rw'\nonumber\\
 &=&-\epsilon r(A\rw')^2-(A\rw')[\epsilon (r A)\rw'+\rw(1-\rw^2)]\nonumber\\
& \le & -r\epsilon(A\rw')^2\nonumber
\end{eqnarray}
  This, together with equation~(\ref{vvpeq}) gives $(A\rw')'/(A\rw')\le -\epsilon /r$. 
Thus, $|A\rw'(r)| \le kr^{-\epsilon}$, or $r^\epsilon|A\rw'(r)|\le k$.  As $\epsilon < 1$, this
contradicts the fact that $\lim_{r\searrow 0}A(r)=\infty$.  \hfill $\blacksquare$ \\

All that remains is to establish the existence of $\lim_{r\searrow 0}\rw'(r)$.  To this end, we define, for any solution of 
equations~(\ref{Aeq}) and (\ref{weq}),
\begin{equation}\label{thdef}
\theta(r)=\arctan \frac{\rw'(r)}{\rw(r)}.
\end{equation}
A simple calculation using equation~(\ref{weq}) yields
\begin{equation}\label{thder}
\theta'=\frac{1}{r^2A}[(\rw^2-1)\cos^2 \theta -r \Phi \cos\theta \sin\theta -r^2A \sin^2 \theta].
\end{equation}

Throughout the rest of this section, we assume $\lim_{r\searrow 0}A(r)=\infty$.
\begin{Lma}\label {thlma}
Suppose  there exists an $\hat r>0$ such that $\rw^2<1$ in $(0,\hat r)$.  Then 
there exists an $0<r_0 <\hat r$  such that $\rw' \ne 0$ in $(0,r_0)$. 
\end{Lma}
\noindent
\textbf{Proof}:  The lemma follows once it is shown that for sufficiently small
$r$, $\theta'|_{\theta =0}<0$ and  $\theta'|_{\theta=\pi/4}>0$.  The first inequality follows 
immediately from equation~(\ref{thder}).

At $\theta=\pi/4$,
\begin{equation}\label{thpi4}
\theta'=\frac{1}{2r^2A}[(\rw^2-1)-r+A(r-r^2)+\frac{(\rw^2-1)^2}{r}+\Lambda r^3].
\end{equation}
For $r$ sufficiently small, $2r-3r^2>0$.  If $(1-\rw^2)\ge r$, then 
\begin{eqnarray}
\theta' & > & \frac{1}{2r^2A}[(\rw^2-1)-r+\frac{(\rw^2-1)^2}{r}+3(r-r^2)] \nonumber \\
 & = & \frac{1}{2r^2A}[(1-\rw^2)(\frac{1-\rw^2}{r}-1)+2r-3r^2] \ge 0 \nonumber 
\end{eqnarray}
so long as $r$ is sufficiently small so that $A(r)>3$.
If $(1-\rw^2)<r$, then
\[\theta'  >  \frac{1}{2rA}[A(1-r)-2]>0\]
so long as $r<1/2$ is sufficiently small so that $A(r)>2$.  \hfill $\blacksquare$    
\begin{Lma}\label{Ainfwp0}
$\lim_{r\searrow 0}\rw'(r)=\rw'_0$ exists.
\end{Lma}
\noindent
\textbf{Proof}:  We choose an $\tilde r>0$ be such that $\Phi(\tilde r)<0$ for all $ 0<r_0<\tilde r$ and, considering \textit{Fact}~\ref{wconst}, an $r_0 \in (0,\tilde r)$ such that either $\rw^2(r_0)\ne 1$ or $\rw^2(r_0)\ne 0$.  Equation~(\ref{weq}) implies (as in \textit{Fact}~\ref{permexit}) $r_0$ can be chosen so that there are only two possibilities:
\begin{list}{(\arabic{prop}):}
{\usecounter{prop}
\lindent}
\item\label{w2>1}for all $0<r<r_0,\;\rw^2(r)>1$ or
\item\label{w2<1}for all $0<r<r_0,\;\rw^2(r)<1$.
\end{list}
\textit{Case~\ref{w2>1}}.  Without loss of generality, we let $\rw>1$.  Equation~(\ref{weq}) and \textit{Fact}~\ref{wconst} reduce everything (by choosing a smaller $r_0$ if necessary) to two subcases:
\begin{list}{(\alph{prop}):}
{\usecounter{prop}
\lindent}
\item\label{negrw'} $\rw'(r)\le 0$ for all $r\in (0,r_0)$ and
\item\label{posw'}$\rw'(r)>0$ for all $r\in (0,r_0)$
\end{list}
\textit{Case\ref{w2>1}a}.  $\lim_{r\searrow 0}\rw(r)$ exists and exceeds 
$1$.  We assume that
$\lim_{r\searrow 0}\rw'(r)$ does not exist and will arrive at a contradiction.  With this assumption, there exists an $\tilde r$ arbitrarily close to $0$ with
$\rw(\tilde
r)>1$, $\rw'(\tilde r)<0$, $\rw''(\tilde r)=0$ and $\rw'''(\tilde r)\le 0$.  Differentiating 
equation~(\ref{weq}) gives
\begin{equation}\label{wdif}
(r^2A)'\rw'' + (r^2A)\rw''' +(r\Phi)\rw'' +[(r\Phi)' +1-3\rw^2]\rw' =0.
\end{equation}
Equation~(\ref{phidef}) gives
\begin{equation}\label{rphider}
(r\Phi)'=2A{\rw'}^2+\frac{2(\rw^2-1)^2}{r^2}-\frac{4\rw\rw'(\rw^2-1)}{r}-2\Lambda r^2.
\end{equation}

For any $\bar \rw>1$, there exists an $\bar r$ such that  $(\rw^2-1)^2/r^2-2\Lambda r^2+1-3\rw^2>0$ whenever $0<r<\bar r$ and
$\rw>\bar
\rw$.  We choose $1<\bar\rw<\lim_{r\searrow 0}\rw(r)$.  It follows from this inequality and equation~(\ref{rphider}) that $\tilde r$ can be chosen so that $[(r\Phi)'
(\tilde r) +1-3\rw^2(\tilde r)]\rw'(\tilde r)<0$.  The left side of equation~(\ref{wdif}), evaluated at $\tilde r$ is negative; a contradiction.\\ \\
\textit{Case~\ref{w2>1}b}.  Equation~(\ref{weq}) precludes a non-positive $\rw''$ in $(0,r)$.  Therefore,
$\lim_{r\searrow 0}\rw'(r)$
exists and is both finite and non-negative.\\ \\
\textit{Case~\ref{w2<1}}.  Lemma~\ref{thlma} implies that $\rw'$ has only one sign.  Therefore,  $\lim_{r\searrow 0}\rw(r)$ must exist.  We take $r$ sufficiently small so that $\rw' \ne 0$ in $(0,r)$.  Now, in this interval, $\rw\rw'(s)=0$ if and only if $\rw(s)=0$ in which case $(\rw\rw')'(s)={\rw'}^2(s)>0$.  It follows that $\rw\rw'$ has only one sign in $(0,\hat r)$ for some $\hat r\in(0,r)$.

If we assume $\lf_{r\searrow 0}\rw'(r)<\lp_{r\searrow 0}\rw'(r)$, there then exists a sequence $\{r_n\}\searrow 0$ with 
$\rw''(r_n)=0$.  Upon multiplying equation~(\ref{weq}) by $\rw'$, \textit{Fact}~\ref{wconst} and $\Phi<0$ give $\rw\rw'(r_n)>0$.  It follows that $\rw\rw'>0$ in $(0,\hat r)$.  Now, equation~(\ref{weq}) multiplied by $\rw$ and evaluated at any nonzero $\bar\rw'\in(\lf_{r\searrow 0}\rw',\lp_{r\searrow 0}\rw')$, gives $(\rw\rw'')_{|\rw'(r)=\bar\rw'}>0$ provided $r$ is sufficiently small.  This contradicts the fact that $\rw'$ has no limit. \hfill $\blacksquare$
\nopagebreak

\end{sloppypar}
\nopagebreak
\bibliographystyle{plain}
\bibliography{b1}

\nocite{pB94}
\nocite{pH73}
\nocite{jS93}
\nocite{jS94}
\nocite{jS951}
\nocite{jS952}
\nocite{jS97}
\nocite{jS98}

\end{document}